\documentclass[vecphys]{svmult}		% vectors in bold face
\usepackage{makeidx}         % allows index generation
\usepackage{graphicx}        % standard LaTeX graphics tool
\usepackage{epsfig}          % another way to include figures
\usepackage{multicol}        % used for the two-column index
\usepackage[bottom]{footmisc}% places footnotes at page bottom
\usepackage{amsmath,amssymb}
\usepackage{cite}

\begin{document}
%%%%%%%%%%%%%%%%%%%%%%%%%%%%%%%%%%%%%%%%%%%%%%%%%%%%%%%%%%%%%%%%%%%%%

\def\jcmindex#1{\index{#1}}
\def\myidxeffect#1{{\bf\large #1}}

% Title
\title*{Collective Coordinate Methods and Their Applicability to $\varphi^4$ Models}
\titlerunning{Collective Coordinate Methods}
\author{Herbert Weigel}

\institute{
Institute for Theoretical Physics, Physics Department, Stellenbosch University, South Africa\\
\texttt{weigel@sun.ac.za}
}
\maketitle
\abstract
{
Collective coordinate methods are frequently applied to study dynamical properties 
of solitons. These methods simplify the field equations - typically partial 
differential equations - to ordinary differential equations for selected 
excitations. More importantly though, collective coordinates provide a 
practical means to focus on particular modes of otherwise complicated dynamical
processes. We review the application of collective coordinate methods in the analysis 
of the kink-antikink interaction within the $\varphi^4$ soliton model and illuminate 
discrepancies between these methods and the exact results from the field equations. 
}

%%%%%%%%%%%%%%%%%%%%%%%%%%%%%%%%%%%%%%%%%%%%%%%%%%%%%%%%%%%%%%%%%%%%%%%%%%%%%%%%%%%
%%%%%%%%%%%%%%%%%%%%%%%%%%%%%%%%%%%%%%%%%%%%%%%%%%%%%%%%%%%%%%%%%%%%%%%%%%%%%%%%%%%
\section{Motivation}

The $\varphi^4$ model represents the simplest non-linear extension of the Klein-Gordon
theory in one time and one space dimensions that contains solutions with localized energy 
densities. These solutions are usually called {\it solitons} though more precisely
they should be named solitary waves~\cite{Ra82,Kivshar:1989ue}. 

Solitons have applications in almost all disciplines of physics. For example they
characterize domains that may emerge in cosmology~\cite{Vachaspati:2006zz,Vilenkin:2000jqa}
or in condensed matter~\cite{Ivanov:1992aa,Trullinger:1976aa}. The localized energy 
densities assign particle structures to solitons and picturing baryons as chiral 
solitons successfully describes many of their static and dynamic 
properties~\cite{Weigel:2008zz}. Solitons typically emerge in field theories
with degenerate vacua such that the soliton configuration takes different vacuum 
values in distinct regimes of spatial infinity. When it takes an infinite amount
of energy to (continuously) transform these vacuum configurations into another,
the so-constructed solitons are named {\it topological} \cite{Manton:2004tk}.

The Lagrange density of the $\varphi^4$ model in one space and one time dimensions 
may be written as compactly as
\begin{equation}
\mathcal{L}=\frac{1}{2}\left[\dot{\varphi}^2-\varphi^{\prime2}\right]
-\frac{1}{2}\left(\varphi^2-1\right)^2\,,
\label{eq:deflag}
\end{equation}
where dots and primes denote (partial) derivatives with respect to time ($t$) and 
space ($x$) coordinates, respectively. Fields and variables have been scaled such 
that the self-interaction strength occurs as an overall constant factor\footnote{Here 
we restrain to the classical description, {\it i.e.} canonical commutation relations 
are not imposed. Then this factor has no dynamical relevance.} which is not 
displayed. The degenerate vacua are simply $\varphi_{\rm vac}=\pm1$ and 
there are two options to connect them thereby defining kink and antikink
solutions. With the above conventions these are
\begin{equation}
\varphi_{K,\overline{K}}(x)=\pm{\rm tanh}(x)\,.
\label{eq:kink4}
\end{equation}
The $\varphi^4$ model (anti)kink is the prototype soliton that illuminates many of the 
structures to be expected for the technically more challenging applications mentioned
above.

Given that these are solutions within a non-linear model, any superposition of two 
solutions will, in general, not be a solution anymore. However, a superposition of two widely 
separated solitons is still approximately a solution as interference contributions to the 
energy vanish. As the separation is reduced practically adiabatically, the super-imposed 
configurations will possess some interaction energy. Within the particle interpretation of 
the soliton, the separation is the distance between two particles and the interaction 
energy becomes the inter-particle potential~\cite{VinhMau:1984jsh}.

In general this interaction can be studied by solving the time dependent field equations 
as an initial value problem. Definitely, for models in one space dimension this in an 
option. But in higher, in particular three, dimensions such computations become 
increasingly demanding. In that case a possible strategy is the introduction of
so-called {\it collective coordinates} that parameterize certain modes of the field 
excitations and reduce the complexity of the field equations drastically. Their 
introduction, to some extent, is a matter of good guess. We are in the lucky situation 
that for soliton models in one space dimension we can compare the solutions to the
exact field equations to those from the reduced equations for the collective
coordinates. This comparison will then determine the quality of the guess. Besides 
these technical advantages, a maybe even more important feature of collective 
coordinates is that they enable us to focus on special excitations of the soliton. 
Studying collective coordinates thus not only sheds light on their dynamical 
relevance but ultimately identifies modes which trigger certain processes thereby
deepening our understanding of the particular soliton model. Here we will 
review this investigation for the kink-antikink system of the $\varphi^4$ model. 

In the following section we will discuss the dynamics of the kink-antikink
interaction resulting from the full field equations. In section~3 we will introduce
collective coordinates to (eventually) detail the dynamics of this interaction.
In section~4 we will briefly reflect on previous approximations to the collective
coordinate approach and discuss one out of many possible generalizations in section~5.
Section~6 shows how the collective coordinate approach is applied to the $\phi^6$ model.
We conclude in section~7 in which we also discuss reasons why the (specific) collective
coordinate approach fails to properly describe the kink-antikink interaction.

\section{Kink-antikink scattering}

Kink-antikink scattering can be simulated as a solution to the field 
equations with suitable initial conditions. Here we discuss this treatment
and its results briefly.

\subsection{Dynamical kink-antikink interaction}

The field equation emerging from the Lagrangian, Eq.~(\ref{eq:deflag}) reads
\begin{equation}
\ddot{\varphi}-\varphi^{\prime\prime}=2\varphi\left(1-\varphi^2\right)\,.
\label{eq:PDE}
\end{equation}
This field equation is a partial differential equation (PDE) and has a unique solution 
when the initial field configuration and its time derivative (velocity) are prescribed. 
The initial conditions suitable to describe the kink-antikink interaction read
\begin{align}
\varphi(x,0)&=\varphi_{\overline{K}}\left(\frac{x}{\sqrt{1-v^2}}-X_0\right)
+\varphi_K\left(\frac{x}{\sqrt{1-v^2}}+X_0\right)-1\,,\cr
\dot{\varphi}(x,0)&=\frac{v}{\sqrt{1-v^2}}\left[
\varphi^\prime_{\overline{K}}\left(\frac{x}{\sqrt{1-v^2}}-X_0\right)
-\varphi^\prime_K\left(\frac{x}{\sqrt{1-v^2}}+X_0\right) \right]\,,
\label{eq:in4}
\end{align}
where $\varphi^\prime_{K,\overline{K}}(x)={\rm sech}^2(x)$. While $v$ is the relative
velocity between the approaching (anti)kinks, $X_0$ measures the initial separation up 
to a factor $\sqrt{1-v^2}$. We take $X_0$ large, so that initially we have a widely 
separated kink-antikink system and interference effects are absent. Then the actual
value of $X_0$ is irrelevant.
Initially kink and antikink approach each other and interact by energy exchange when 
they are close enough. Eventually energy is transferred to fluctuations about the 
kink-antikink system. This energy is then not available as translational energy and 
the components do not separate. Only when the fluctuations eventually release that 
energy again, do the kink and antikink depart. This produces so-called {\it bouncing} or 
{\it resonant} solutions in which kink and antikink structures partially separate 
but then {\it turn around} and approach again. The nature of the solutions varies 
with the initial velocity $v$.

\subsection{Bouncing solutions and bounce windows}

\begin{figure}
\bigskip
\centerline{
\epsfig{file=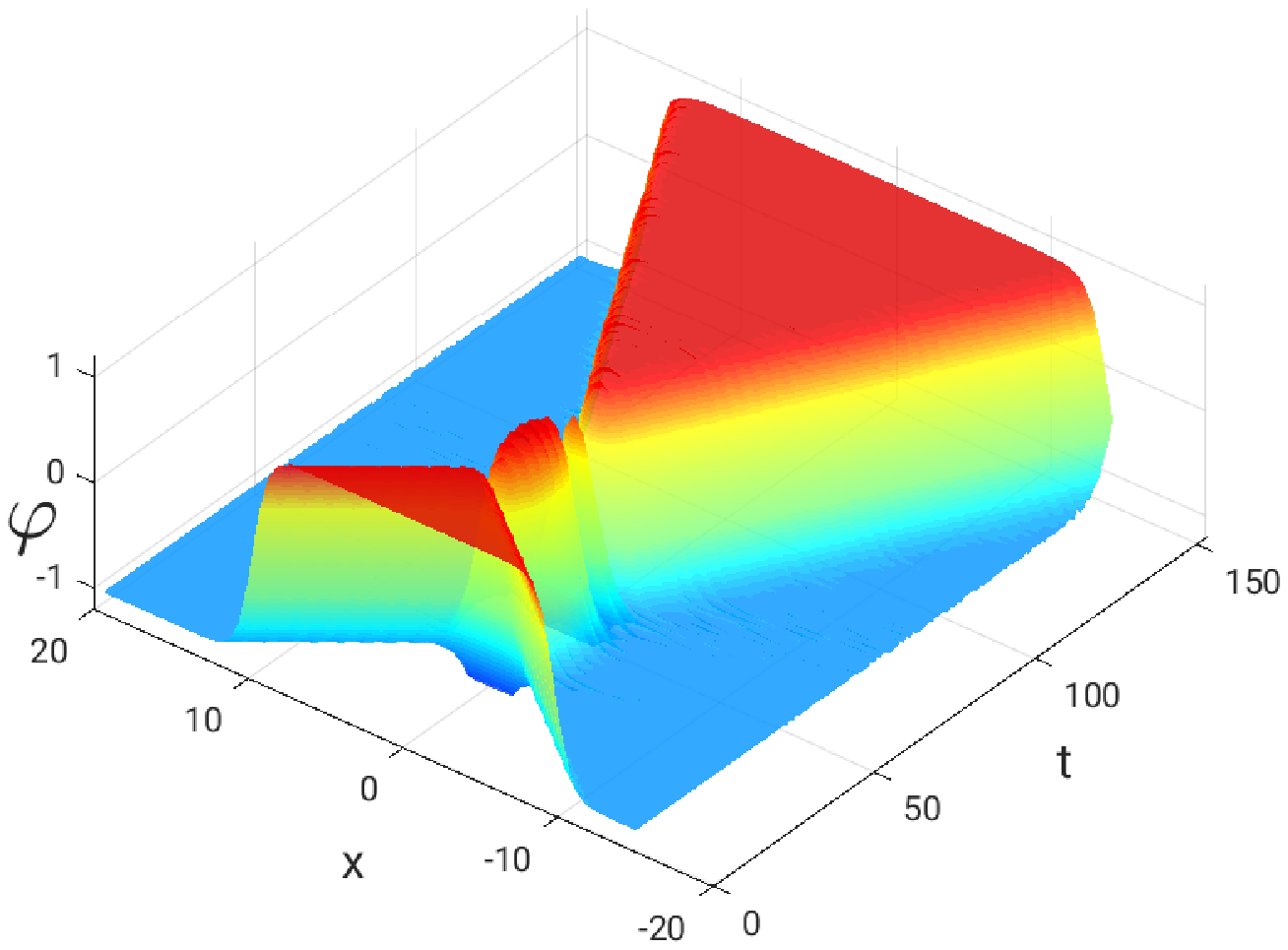,width=6cm,height=5cm}\hspace{0.1cm}
\epsfig{file=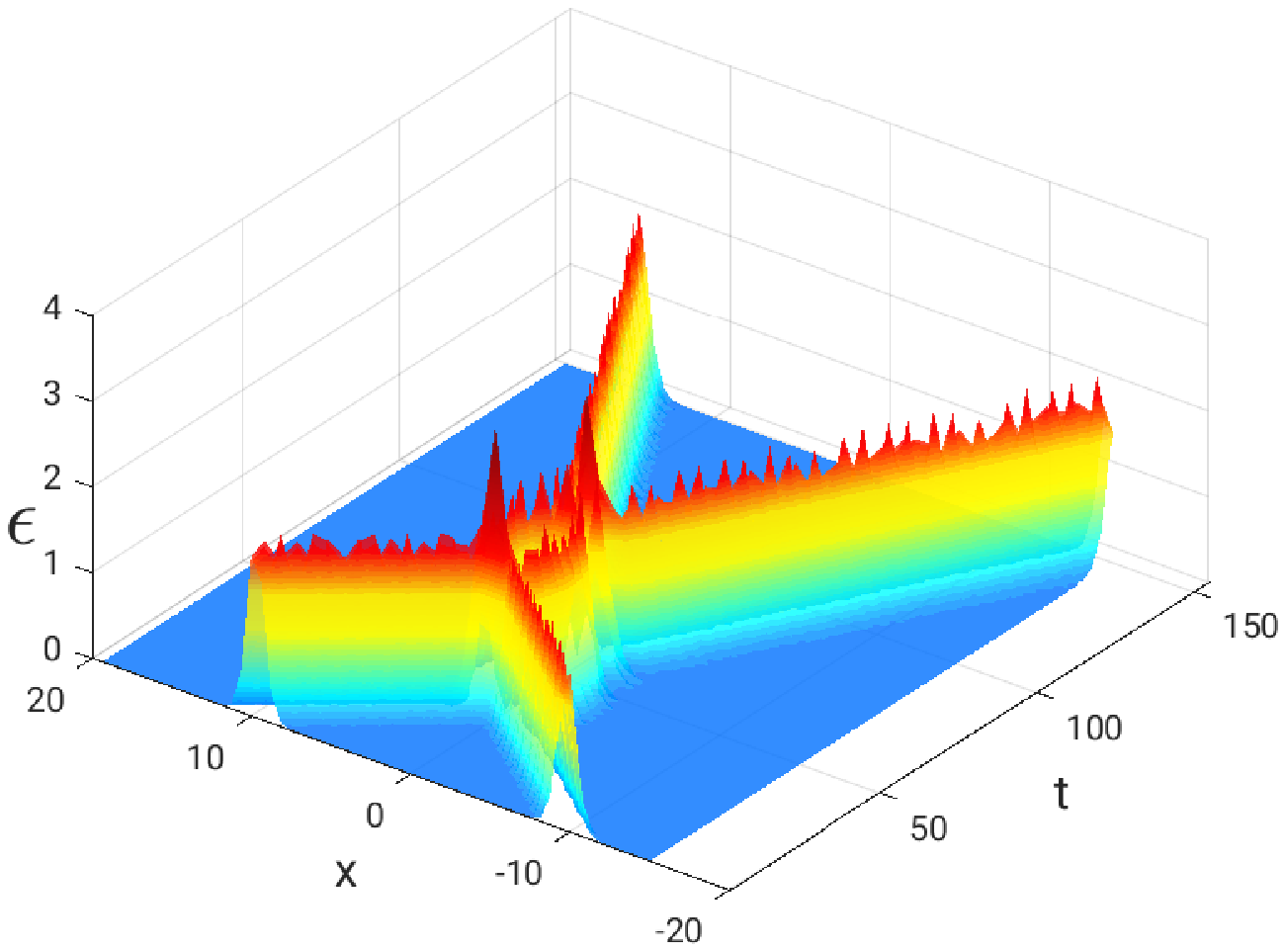,width=6cm,height=5cm}}
\caption{\label{fig:bounce}(Color online) Typical solution to the initial value problem
defined through equations (\ref{eq:in4}) and (\ref{eq:PDE}) for $v=0.251$ for the
field $\varphi$ in the left panel and the energy density $\epsilon$ from Eq.~(\ref{eq:edens})
in the right panel.}
\bigskip
\end{figure}

A typical solution to the above described initial value problem is shown in 
figure \ref{fig:bounce}.
This example exhibits the main features of the solutions to the full field equations.
Kink and antikink approach each other, and rather than immediately departing to spatial
infinity, they bounce and interact again before finally separating. Depending on the 
initial velocity several bounces may occur. 

The bouncing solutions are most clearly discussed by defining a separation variable
of the kink and antikink structure observed in the numerical simulation. We identify 
this to be the position of the antikink and we extract it as the expectation
value of the coordinate along the positive half-line \cite{Weigel:2013kwa}
\begin{equation}
\langle x \rangle_t=\frac{\int_0^\infty dx\, x\, \epsilon(t,x)}
{\int_0^\infty dx\,  \epsilon(t,x)}\,.
\label{eq:xn}
\end{equation}
Here
\begin{equation}
\epsilon(t,x)=\frac{1}{2}\left[\ddot{\varphi}+\varphi^{\prime\prime}
+\left(\varphi^2-1\right)^2\right]
\label{eq:edens}
\end{equation}
is the classical energy density from the time dependent numerical solution to 
Eq.~(\ref{eq:PDE}) with initial conditions given in Eq.~(\ref{eq:in4}). 

\begin{figure}[t]
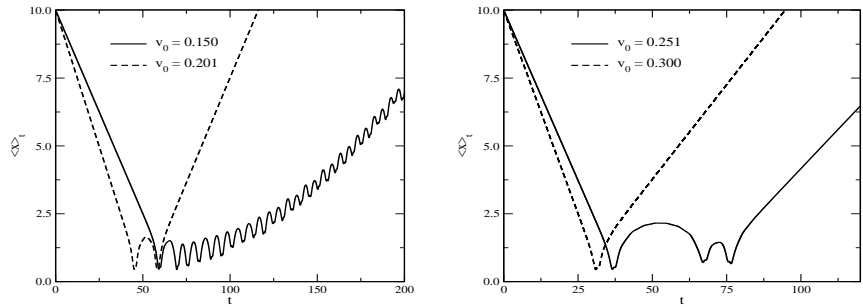

\bigskip
\centerline{
\epsfig{file=PDE1.eps,width=5.4cm,height=4cm}~~~~
\epsfig{file=PDE2.eps,width=5.4cm,height=4cm}}
\caption{\label{fig:pde1}Separation between kink and antikink extracted from the 
solutions to the partial differential equation~(\ref{eq:PDE}) according to 
Eqs.~(\ref{eq:xn}) and~(\ref{eq:edens}) for different initial velocities.}
\bigskip
\end{figure}

Typical results for $\langle x \rangle_t$ are shown in figure \ref{fig:pde1}. 
Essentially three types of structures are observed. At low initial velocity, 
kink and antikink approach and bounce several times before they eventually separate. 
Even after separation the relative fluctuations diminish only gradually and the 
distinct kink and antikink configurations can be observed only for very late times.
At moderate velocities, the kink-antikink system bounces a few times and then separates 
at slightly lower velocity. That is, at late times some of the energy remains in
fluctuation modes. At large velocities the system separates without any bounce, but 
still some energy is contained in fluctuations about the individual (anti)kink as can
be seen from the final velocity being smaller than the initial. It is obvious
that at large relative velocities there is always sufficient energy available for 
immediate separation. 

These bounce structures of the kink-antikink interaction have been exhaustively discussed 
in Refs. \cite{Moshir:1981ja,Campbell:1983xu,Belova:1988fg,Anninos:1991un,Goodman:2005} and 
reviewed in Ref. \cite{Belova:1997bq}. In particular the critical velocity above which no 
bounces occur has been extracted to be $v_{\rm c}=0.260$ \cite{Moshir:1981ja,Campbell:1983xu}. 
Also certain intervals for the initial velocity below $v_{\rm c}$, so-called bounce windows, 
have been identified in which zero, one, two or any larger number of bounces occurs before 
the final separation.

\section{Collective coordinates}

We want to gain a deeper understanding of the rich bounce structures observed in the 
solutions to the partial differential equations. It is particularly challenging 
to identify those modes in which the energy is temporarily stored preventing the 
kink and antikink to separate immediately. For this purpose we want to model those
solutions by admitting only particular modes. The introduction of collective 
coordinate is a very promising technique for this endeavor.

\subsection{Large vs. small amplitude fluctuations}

There are two distinct types of fluctuations about classical localized configurations in a field 
theory. In the standard procedure small amplitude fluctuations, $\eta(x,t)$ are introduced (for 
quantization) \cite{Ra82}. In a first step the field is parameterized as
$$
\varphi(x,t)=\varphi_{\rm cl}(x)+\eta(x,t)\,,
$$ 
where $\varphi_{\rm cl}(x)$ represents the classical configuration, as {\it e.g.} the kink in 
Eq.~(\ref{eq:kink4}). In the second step, the Lagrangian is expanded in powers of $\eta$. 
The linear order vanishes when $\varphi_{\rm cl}(x)$ solves the field equation and contributions
beyond the harmonic order are omitted. This harmonic expansion is rigorous in the $\hbar$ 
counting when $\eta(x,t)$ is normalized by the canonical commutation relations and produces 
the leading quantum corrections to properties of the classical configuration, most prominently 
the vacuum polarization (or Casimir) energy \cite{Graham:2009zz}.

The spectrum of the small amplitude fluctuations is particularly interesting in the $\varphi^4$ 
model. Besides the translational zero mode (see below) there is a bound state below threshold 
(at frequency $\omega_{\rm th}=2$, using dimensionless variables),
\begin{equation}
\eta(x,t)={\rm e}^{-i\sqrt{3}t}\chi(x)
\qquad {\rm with} \qquad 
\chi(x)=\frac{{\rm sinh}(x)}{{\rm cosh}^2(x)}\,.
\label{eq:shape}
\end{equation}
In the context of the kink-antikink interaction it has been conjectured that during bounces
most of the energy is stored in this so-called {\it shape mode} \cite{Goodman:2005}.

However, often there are also non-harmonic modes, that is, modes that are not subject to 
a restoring force and thus may acquire large amplitudes.  Those modes are typically parameterized 
by time dependent variables. These independent variables are called collective coordinates as 
they describe collective motions of the classical field configuration.

\subsection{Identification of crucial modes}

As discussed above, a number of features motivates the introduction of collective coordinates.
Unfortunately, a suitable choice may be a matter of good guess. Since the collective modes are
supposed not to experience any or only a small restoring force the (would-be) zero modes of
the classical configurations are good candidates. The excitation of these modes does not alter
the energy. The top candidate is, of course, the translation of the localized configuration which
generates its linear momentum. As a further advantage, the introduction of collective 
coordinates simplifies an intricate quantum field theory problem into one of quantum mechanical 
nature. In higher dimensions rotations (both in coordinate and internal spaces) generate good 
quantum numbers that are not possessed by the classical soliton. A prominent example is the 
Skyrme model \cite{Skyrme:1961vq} in which canonical quantization of the collective coordinates 
generates baryon states \cite{Adkins:1983ya}. (For a review see Ref. \cite{Weigel:2008zz}.)
This procedure can be extended to the case of approximate, or (would-be) zero modes with the
symmetry violation treated in perturbation theory.

\subsection{Separation as collective coordinate}
\label{ssec:separation}

For the discussion of the interaction between kink and antikink it is obvious to introduce 
a time dependent collective coordinate, $X(t)$, for their separation. It takes kink and antikink
with opposite velocities and parameterizes the field configuration as
\begin{equation}
\varphi_c(x,t)=\varphi_K(\xi_{+})+\varphi_{\overline{K}}(\xi_{-})-1\,.
\label{eq:ccsep}\end{equation}
where
$\xi_\pm=\xi_\pm(x,t)=\frac{x}{\sqrt{1-v^2}}\pm X(t)$. Here $v$ equals the relative velocity 
in the initial conditions, Eq.~(\ref{eq:in4}). Substituting this parameterization and 
integrating over the spatial coordinate $x$ produces the Lagrange function
\begin{equation}
L(X,\dot{X})=a_1(X)\dot{X}^2-a_2(X)\,,
\label{eq:col1}\end{equation}
with mass ($a_1$) and potential ($a_2$) functions
\begin{align}
a_1(X)&=\frac{1}{2}\int_{-\infty}^\infty dx\,
\left[\varphi^\prime_K(\xi_{+})+\varphi^\prime_{\overline{K}}(\xi_{-})\right]^2\cr
a_2(X)&=\frac{1}{2}\frac{1}{1-v^2}\int_{-\infty}^\infty dx\,
\left[\varphi^\prime_K(\xi_{+})-\varphi^\prime_{\overline{K}}(\xi_{-})\right]^2
+\frac{1}{2}\int_{-\infty}^\infty dx\,
\left[\varphi_c(x,t)-1\right]^2\,,\quad
\label{eq:colint1}\end{align}
that parametrically depend on the collective coordinate $X$ via $\xi_{\pm}$. These integrals 
can be computed analytically \cite{Kudryavtsev:1075ku,Sugiyama:1979mi,Campbell:1983xu} and 
an example is detailed in the appendix of Ref. \cite{Takyi:2016tnc}. It is, however, equally 
efficient to calculate them numerically since the equation of motion for the collective 
coordinate 
\begin{equation}
\ddot{X}=-\frac{1}{2a_1(X)}\left[\frac{da_1(X)}{dX}\,\dot{X}^2+\frac{da_2(X)}{dX}\right]
\label{eq:ODE}\end{equation}
must also be solved numerically. Another motivation to obtain these integrals numerically is
that small variations of the parameterization, Eq.~(\ref{eq:ccsep}) can then be easily 
accommodated. Since the time dependence of the field configuration only comes 
via $X(t)$ this is an ordinary differential equation (ODE) and thus technically less 
challenging than the full field equations (\ref{eq:PDE}).

\begin{figure}[b]
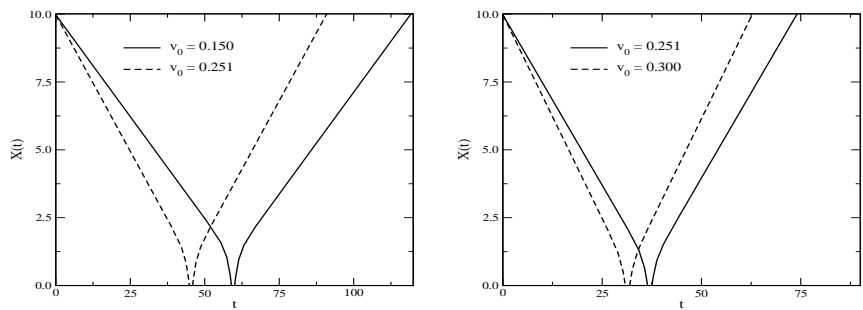

\bigskip
\centerline{
\epsfig{file=COL1.eps,width=5.4cm,height=4cm}~~~~
\epsfig{file=COL2.eps,width=5.4cm,height=4cm}}
\caption{\label{fig:col1}Solutions to the equation of motion resulting from the
collective coordinate Lagrangian, Eq.~(\ref{eq:col1}).}
\bigskip
\end{figure}

Solutions with initial velocities analog to those in equation (\ref{eq:in4})
\begin{equation}
X(0)=10 \qquad {\rm and}\qquad \dot{X}(0)=\frac{v}{\sqrt{1-v^2}}
\label{eq:initial}\end{equation}
are displayed in figure \ref{fig:col1}.
Obviously these solutions do not produce any bounce structures. This is not surprising 
as there is only a single degree of freedom that cannot exchange energy with any other 
mode. The prime candidate mode to include for this energy exchange is the shape mode, 
Eq.~(\ref{eq:shape}) whose incorporation we will discuss next.

\subsection{Excitation of shape mode}

We now go one step further and include the amplitude of the shape mode as a collective
coordinate \cite{Sugiyama:1979mi,Takyi:2016tnc}
\begin{equation}
\varphi_c(x,t)=\varphi_K(\xi_{+})+\varphi_{\overline{K}}(\xi_{-})-1
+\sqrt{\frac{3}{2}}\left[A(t)\chi(\xi_{-})+B(t)\chi(\xi_{+})\right]\,\,\,.
\label{eq:cc4}
\end{equation}
Note that both the kink and the antikink are accompanied by shape modes. Again
this parameterization is substituted into the Lagrangian and integration over
the coordinate $x$ yields the Lagrange function
\begin{align}
L(X,\dot{X},A,\dot{A},B,\dot{B})&=a_1(X)\dot{X}^2-a_2(X)+
a_3\left[\dot{A}^2+\dot{B}^2\right]-a_4\left[A^2+B^2\right]\cr
&\hspace{1cm}
+\overline{a}_3\dot{A}\dot{B}-\overline{a}_4AB
+a_5(X)\left[A-B\right]+\ldots\,.
\label{eq:col2}\end{align}
Only a few of the many terms that exhibit the essential structures have been displayed. 
The full Lagrange function as well as explicit expressions for the coefficient functions 
are listed in Ref. \cite{Takyi:MSC2016}. 

Complicated Euler Lagrange equations that generalize Eq.~(\ref{eq:ODE}) are 
derived from Eq.~(\ref{eq:col2}). At $t=0$ kink and antikink approach each other while 
the shape modes are not excited initially, {\it i.e.}
$A(0)=B(0)=0$ and $\dot{A}(0)=\dot{B}(0)=0$. In general the amplitudes $A$ and $B$ 
are independent. However, the above listed terms, in particular the linear term
involving $a_5(X)$, suggest (and that is found for the full Lagrange function) that 
if $A(t)$ is a solution, so is $B(t)\equiv-A(t)$. That is, initial conditions with 
$B(0)=-A(0)$ and $\dot{B}(0)=-\dot{A}(0)$ produce $B(t)\equiv-A(t)$. In case there 
is no shape mode initially it is thus sufficient to only consider $A(t)$. From now on 
we will therefore adopt that specific collective coordinate configuration. Then
the shape mode part in the collective coordinate parameterization 
becomes \cite{Sugiyama:1979mi}
\begin{equation}
\sqrt{\frac{3}{2}}\left[A(t)\chi(\xi_{-})+B(t)\chi(\xi_{+})\right]
\quad \longrightarrow\quad
\sqrt{\frac{3}{2}}\left[\chi(\xi_{-})-\chi(\xi_{+})\right]A(t)\,.
\label{eq:ccSug}\end{equation}
As $X\to0$ we have $\xi_{-}\to\xi_{+}$ and the coefficient of $A(t)$ vanishes.
This leads to a null-vector singularity \cite{Caputo:1991cv,Goodman:2007bb} as 
the amplitude is not well defined in that limit and causes a major obstacle to
introducing the amplitude of the shape mode as a collective coordinate. This 
obstacle can only be circumvented by particular approximations or modifications.

The comparison with the (integrable) sine-Gordon soliton further motivates to 
consider the shape mode as the prime candidate for intermediate energy storage. 
That model actually lacks a shape mode type solution as a small amplitude 
fluctuation off the soliton. Neither do bounces occur in the soliton-antisoliton
interaction as the known analytic expression \cite{Ra82} for the time-dependent 
system reveals; only some time delay or phase shift is observed in the 
soliton-antisoliton interaction \cite{Campbell:1983xu}. With generalizations or 
additions of impurities, bounces emerge \cite{Peyrard:2983ef} that have been 
analyzed using collective coordinates in Refs. \cite{Fei:1992ab,Goodman:2004cd}.

\subsection{Approximations in collective coordinate calculations}

\begin{figure}[t]
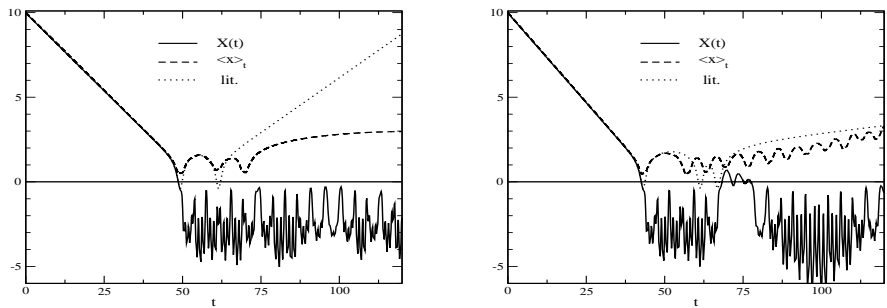

\bigskip
\centerline{
\epsfig{file=HARM1.eps,width=5.3cm,height=4cm}\hspace{1cm}
\epsfig{file=HARM2.eps,width=5.3cm,height=4cm}}
\caption{\label{fig:HARM1}Kink-antikink separation as function of time. Collective 
coordinate calculation of Eqs.~(\ref{eq:cc4}) and~(\ref{eq:ccSug}): full lines; 
field equations: dashed lines; collective coordinate approach from 
Refs. \cite{Anninos:1991un,Goodman:2005}: dotted lines. 
Left panel: $v=0.184$; right panel $v=0.212$.}
\bigskip
\end{figure}

Early approaches to solve the equations of motion for the collective
coordinates $X(t)$ and $A(t)$ circumvented the null-vector problem by omitting 
the $X$ dependences in $a_3$ and $a_4$ by equating them to their constant values
from the respective $X\to\infty$ limits. Essentially this omits the interference 
between the two shape modes at $\xi_{\pm}$. Furthermore non-harmonic terms
in $A$, {\it i.e.} those beyond quadratic order in the Lagrangian were
discarded \cite{Anninos:1991un,Goodman:2005,Goodman:2007bb,Goodman:2015ina}.
Calculations (denoted 'lit' in figure \ref{fig:HARM1}) based on these
approximations indeed showed surprising agreement with the results from
the full field equation (\ref{eq:PDE}). Unfortunately those calculations 
took $a_5(X)$ from the original paper \cite{Sugiyama:1979mi} which suffers
from a misprint that made its way through the literature.
Correcting\footnote{Remarkably, the fitted function $a_5(X)\propto {\rm e}^{-X}$ 
studied in Refs. \cite{Goodman:2007bb,Goodman:2015ina} reasonably resembles the 
actual behavior.} the misprint \cite{Takyi:2016tnc} and re-analyzing the equations 
of motion using the same approximations removes any agreement with the
results from Eq.~(\ref{eq:PDE}) as displayed in figure \ref{fig:HARM1}.
The separation coordinate turns negative with a significant modulus 
and wild oscillations. Only occasionally it returns to the positive regime 
but essentially it is trapped below zero. Obviously these approximations,
in particular omitting higher powers in $A$, are not consistent with the
results from numerically simulating the full field equations.

\subsection{Orbits of collective coordinates}

\begin{figure}
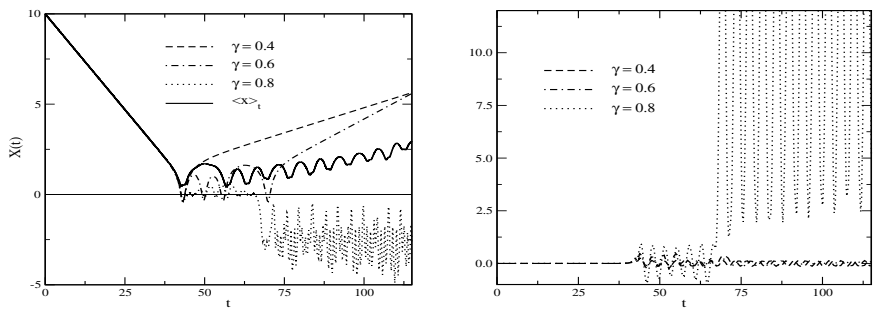

\bigskip
\centerline{
\epsfig{file=Xga.eps,width=5.4cm,height=4cm}\hspace{0.6cm}
\epsfig{file=Aga.eps,width=5.4cm,height=4cm}}
\caption{\label{fig:XAga}Collective coordinates from scaled source,
Eq.~(\ref{eq:XAga}) for $v=0.212$ in the harmonic approximation.}
\bigskip
\end{figure}

Comparing the numerical solutions with (Fig. \ref{fig:HARM1}) and without 
(Fig. \ref{fig:col1}) a collectively excited shape mode with that of 
the full solution in figure \ref{fig:pde1} suggests that this mode may 
indeed be significant, but that its incorporation as the only collective 
coordinate on top of the separation exaggerates its role. To further 
investigate this conjecture we scale its source by a constant $\gamma$:
\begin{equation}
a_5(X)\,\longrightarrow\,\gamma a_5(X)
\label{eq:XAga}\end{equation}
and again solve the equations of motions within the above discussed 
approximations. The resulting time dependences for both the separation $X$
and the shape mode amplitude $A$ are displayed in figure \ref{fig:XAga}.
We see that a moderate rescaling produces trajectories for $X(t)$ that
compare reasonably well with $\langle x\rangle_t$ extracted from the full 
field equations. For small $\gamma$ the bounce structure disappears ({\it cf.} 
Sec. \ref{ssec:separation}) while taking the source with full strength produces 
too large amplitudes for the shape mode and ties the separation coordinate
into the negative domain. However, even for $\gamma=0.5$ we find the critical 
velocity to be about $v_{\rm c}=0.33$ which is significantly larger than the 
value extracted from the exact solution. This suggests that, though the shape 
plays a significant role, other modes seem equally relevant.

Note, however, that these are intermediate conclusions, as the collective 
coordinate approach has been furnished by a number of approximations. We will 
elaborate on these approximations in the following section.

\section{Comments on approximations}

In the previous section we have seen that, in the context of kink-antikink 
scattering, frequently adopted approximations within the collective coordinate 
formulation do not reproduce the results from the field equations. It is 
therefore suggestive that a better agreement will be obtained when these
approximations will be abandoned. 

Most of the approximations for the collective coordinate approach have been 
implemented to gain simplifications that lead to analytic results or to avoid 
technical problems. With a fully numerical approach, some of these approximations
can easily be abandoned.

\subsection{Non-harmonic contributions}

When omitting the non-harmonic contributions of the shape mode amplitude, the 
collective coordinate equations of motion can be utilized to analytically 
predict the critical velocity \cite{Sugiyama:1979mi,Goodman:2005}. One may
also argue for their omission on the basis that the shape mode as bound state
in the background of a single kink results from exactly that approximation.
However, the results shown in figure \ref{fig:XAga} are actually inconsistent 
with the harmonic approximation. It is obvious that the amplitude of the shape 
mode may be large and therefore $\mathcal{O}(A^2)$ terms may not be omitted 
from the equations of motion. Inclusion of these higher order terms 
may hamper the amplitude of the shape mode from becoming large by absorbing 
a significant amount of energy. In turn this might improve the quality of
collective coordinate approach. Stated otherwise, the non-harmonic terms are 
hoped to assist towards reducing the amplitude $A$ such that the small amplitude 
approximation, which introduced the shape mode in the first place, is indeed 
justified.

\subsection{Null-vector singularity}

We have already mentioned that the collective coordinate parameterization
of Eq.~(\ref{eq:ccSug}) produces an ill-defined amplitude of the shape mode 
when $X\to0$ for the solution with $A(t)=-B(t)$ which is realized when the 
initial conditions obey this relation. To show that this holds true even 
without the identification $A\equiv-B$ we consider the kinetic terms 
involving the shape mode. The associated coefficients are 
\begin{equation}
a_3(X)=\frac{3}{4}\int dx \,\chi^2(\xi_{\pm})=
\frac{3}{4}\sqrt{1-v^2}\int dx\, \chi^2(x)
\,\,\, {\rm and}\,\,\,
\overline{a}_3(X)=\frac{3}{2}\int dx\,\chi(\xi_{+})\chi(\xi_{-})\,.
\label{eq:null1}\end{equation}
When $X=0$ we have that $\xi_{+}=\xi_{-}=\frac{x}{\sqrt{1-v^2}}$
so that $\overline{a}_3(0)\to2 a_3(0)$. Then the coefficient matrix 
in the collective coordinate Lagrangian becomes
$$
\begin{pmatrix}
a_3(X) & \overline{a}_3(X) \cr \overline{a}_3(X) & a_3(X)
\end{pmatrix}
\sim a_3(0) \begin{pmatrix} 1 & 1 \cr 1 & 1\end{pmatrix}\,,
$$
whose null-vector $\begin{pmatrix}1 \cr -1\end{pmatrix}$ is indeed the 
solution $A(t)=-B(t)$ \cite{Caputo:1991cv}. The above matrix cannot be 
inverted and it becomes impossible to formulate the equations of motion 
as $\ddot{A}=\ldots$~~and $\ddot{B}=\ldots$. The potential part containing 
$a_4(0)$ and $\overline{a}_4(0)$ has the same null-vector. In the numerical 
approach this singularity allows arbitrarily large amplitudes of the shape 
mode when $X\sim0$. This suggests to modify the collective coordinate 
parameterization such that $X\to0$ is energetically disfavored. We will get 
back to this in the next section.

\subsection{Disagreement with solution to field equations}

\begin{figure}[t]
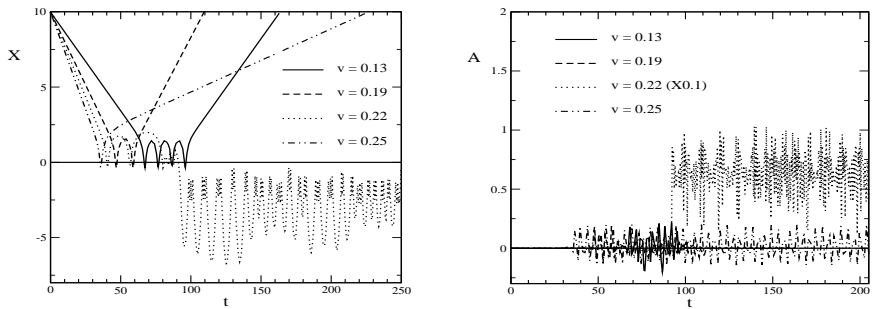

\bigskip
\centerline{
\epsfig{file=xphi.eps,width=5.4cm,height=4cm}\hspace{0.6cm}
\epsfig{file=aphi.eps,width=5.4cm,height=4cm}}
\caption{\label{fig:ODEphi4}Solution to the collective coordinate equations
of motion in the $\varphi^4$ model. 
Left panel: kink-antikink separation, right panel: amplitude of shape mode 
(note the change of scale in the $v=0.22$ entry).}
\bigskip
\end{figure}

In view of the above discussion it is therefore suggestive to solve 
the collective coordinate equations as they emerge from the Lagrange function,
Eq.~(\ref{eq:col2}) for $X(t)$ and $A(t)\,\,(=-B(t))$. At this point the sole 
approximation is to omit $\overline{a}_3(X)$ and $\overline{a}_4(X)$ (that is,
keep the coefficients of the harmonic terms at their $X\to\infty$ values) to 
avoid the technical null-vector problem discussed above. The results of these 
calculations are shown in figure~\ref{fig:ODEphi4}.
It turns out that the energy associated with the source term for the 
shape mode, $E_5=-a_5(X)A$, may absorb much energy when $X$ becomes 
negative and trapping type solutions with large amplitudes of the shape 
mode emerge. Hence the inclusion of the non-harmonic terms does not 
produce an a posteriori justification for implementing the harmonic 
approximation. The entry with $v=0.22$ in figure \ref{fig:ODEphi4} is 
a typical example thereof. 

\begin{figure}
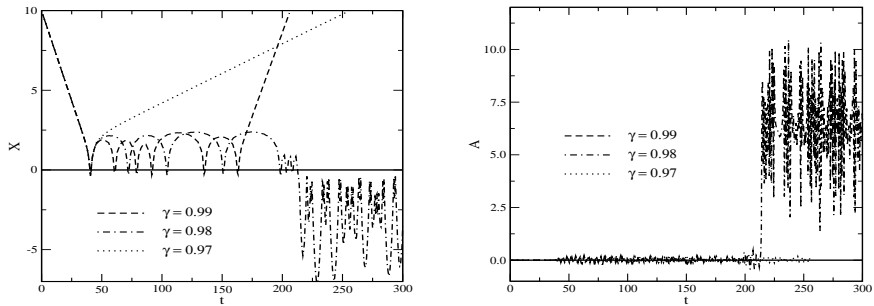

\bigskip
\centerline{
\epsfig{file=Xgafull.eps,width=5.4cm,height=4cm}\hspace{0.6cm}
\epsfig{file=Agafull.eps,width=5.4cm,height=4cm}}
\caption{\label{fig:XAgafull}Collective coordinates from scaled source,
Eq.~(\ref{eq:XAga}) for $v=0.22$.}
\bigskip
\end{figure}

To further reflect on the quality of the collective coordinate formulation we 
again apply the scaling of Eq.~(\ref{eq:XAga}) to one of the worst results to 
this case. The resulting time dependence of the separation variable is shown 
in figure \ref{fig:XAgafull}.
Obviously there is quite a sensitivity with respect to the strength 
of the source term. It is unlikely that the collective coordinate 
approach is sensitive enough to properly account for such fine details. 
This also concerns the predictions for the outgoing velocities.

Most notably, however, we find that without the many approximations 
(and $\gamma=1$), the collective coordinate approach predicts a critical 
velocity of $v_{\rm c}=0.247$ above which trapping or bounce type solution 
cease to exist. This still compares favorably to the result from the full 
field equations of~$0.260$~\cite{Moshir:1981ja,Campbell:1983xu}. Yet, in 
view of the various discrepancies between the two approaches below $v_{\rm c}$, 
this may be a mere artifact.

\section{Modifications}

Obviously negative $X(t)$, in particular with a large modulus, should not
appear\footnote{The comparison with the solutions to the full field equations
proceeds via $\langle x\rangle_t$ which is positive definite, {\it cf.} 
Eq.~(\ref{eq:xn}).  Hence any such comparison is bound to fail for negative~$X(t)$. 
We will therefore also consider the actual field configurations in 
section \ref{sec:map}.}. We have seen that when $X(t)$ is trapped at (large) 
negative values, even with the non-harmonic terms included, the amplitude 
of the shape mode contradicts the assumption for the small amplitude 
approximation which has been the point of departure for the shape mode
and its subsequent use as collective coordinate.

Originally, taking $X$ as a collective coordinate was motivated from the 
configuration containing widely separated kink and antikink structures. 
Of course, such a starting point does not determine a unique collective 
coordinate parameterization as we can always add any contribution that 
vanishes as $X\to\infty$. Here we will explore one such modification.

\subsection{Kink-antikink penetration}

The collective coordinate parameterization of Eq.~(\ref{eq:ccsep}) turns into
a vacuum configuration of zero energy when $X\sim0$. This causes attraction for 
small and moderate separation and eventually produces solutions that are trapped 
with negative $X$ and correspond to (un-desired) configurations in which kink and 
antikink have penetrated each other. We thus modify the collective coordinate 
description and introduce a variational parameter $q$
\begin{equation}
\varphi_c(x,t)=\varphi_K(\xi_{+})+\varphi_{\overline{K}}(\xi_{-})-{\rm tanh}(qX)
+\sqrt{\frac{3}{2}}A(t)\left[\chi(\xi_{-})-\chi(\xi_{+})\right]\,.
\label{eq:cc4mod}
\end{equation}
For $q\gg1$ this agrees with the original formulation, Eq.~(\ref{eq:ccsep}) 
when $X\ge0$ which embraces the desired initial condition. In contrast to 
Eq.~(\ref{eq:ccsep}), however, it is also a 
solution to the field equations when $X\to-\infty$ (with $A=0$). Actually
the configuration, Eq.~(\ref{eq:cc4mod}) is anti-symmetric under $X\to-X$.

\begin{figure}
\bigskip
\centerline{
\epsfig{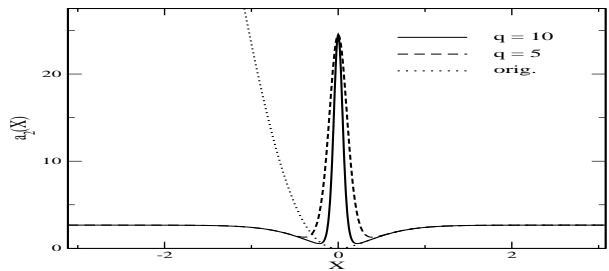}}
\caption{\label{fig:a2pot}Potential for the separation collective coordinate
from the modified parameterization, Eq.~(\ref{eq:cc4mod}) with $v=0.2$. 
The entry 'orig.' refers to the background of Eq.~(\ref{eq:ccsep}).}
\bigskip
\end{figure}

With the ${\rm tanh}(qX)$ term, the $X\sim0$ configuration ceases to be a 
vacuum configuration. In turn this produces a repulsive potential, $a_2(X)$ for 
$X\to0$. When $q$ is large, the repulsion resembles a peak on top of the 
intermediate attraction of the original parameterization. For moderate and 
small $q$ it essentially removes that attraction. This scenario is pictured in
figure \ref{fig:a2pot}.

We note that the parameterization, Eq.~(\ref{eq:cc4mod}) is different 
from the one used in Ref. \cite{Takyi:2016tnc} since here we immediately 
impose equal amplitudes for the two shape modes via Eq.~(\ref{eq:ccSug}). 
Though that equality also results from the equations of motion associated to  
Eq.~(\ref{eq:cc4mod}), various approximation/modifications 
may yield different numerical results (Most of the numerical simulations 
in Ref. \cite{Takyi:2016tnc} assumed the $X\to\infty$ values for the harmonic 
terms.)

There is no fundamental principle in choosing a value for $q$ other that 
Eq.~(\ref{eq:cc4mod}) initially resembles a well separated kink antikink system,
{\it i.e.} $qX_0\gg1$. Later we will see that there is not much variation in
the structure of the soliton with $q$ once it is taken large enough. A guiding 
principle to select $q$ could, for example, be to tune it for any given value 
of $v$ such as to achieve maximal agreement with the solution from the field 
equations. This was done in Ref. \cite{Demirkaya:2017euk}, however in the 
context of the $\phi^6$ model, {\it cf.} section \ref{sec:phi6}.

\subsection{Improved agreement with solution to field equations}

\begin{figure}[t]
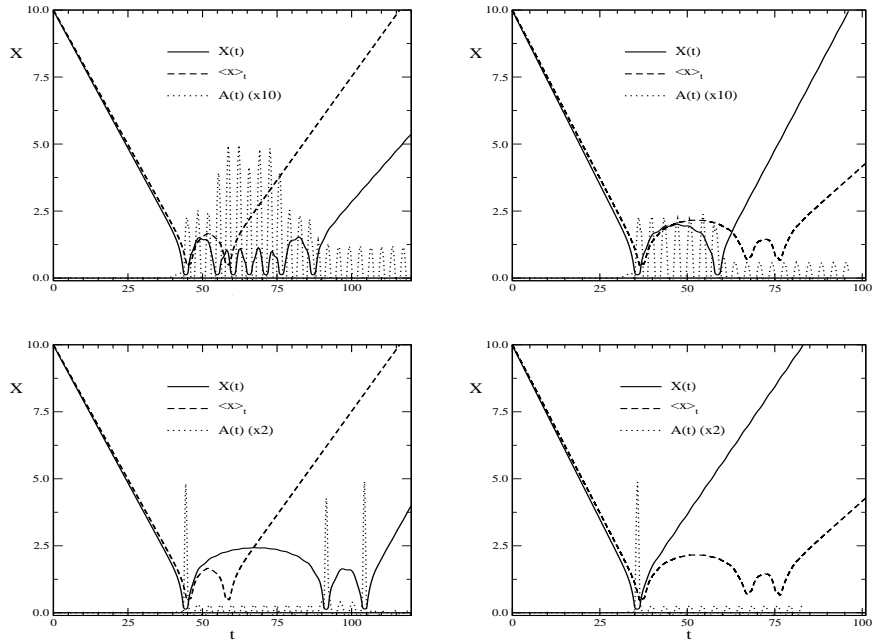

\bigskip
\centerline{
\epsfig{file=XAv201a3a4.eps,width=5.4cm,height=4cm}\hspace{0.6cm}
\epsfig{file=XAv251a3a4.eps,width=5.4cm,height=4cm}}
\bigskip
\centerline{
\epsfig{file=XAv201.eps,width=5.4cm,height=4cm}\hspace{0.6cm}
\epsfig{file=XAv251.eps,width=5.4cm,height=4cm}}
\caption{\label{fig:XA}
Collective coordinate calculations with $q=10$ in Eq.~(\ref{eq:cc4mod})
with all terms from the collective coordinate Lagrangian included. 
Top graphs: Coefficients of harmonic terms kept constant at their 
$X\to\infty$ values. Bottom graphs: $X$ dependence of harmonic terms included.
Left panel: $v=0.201$, right panel: $v=0.251$. Note the different scales
for $A(t)$.}
\bigskip
\end{figure}

We discuss the numerical solutions for the collective coordinate description 
of Eq.~(\ref{eq:cc4mod}) and compare typical results with those from the full 
calculation in figure \ref{fig:XA}. The two graphs on top originate from taking 
the coefficients of the quadratic terms, both for the kinetic~($a_3$) and the 
potential~($a_4$) constants at their $X\to\infty$ values. This approximation is 
adapted from earlier attempts to avoid the null-vector problem. We see that 
this collective coordinate approach indeed describes a finite number of bounces 
and that these bounces come together with measurable excitations of the shape 
mode. However, neither the number of bounces nor the time scale at which these 
bounces occur are properly reproduced by the collective coordinate calculation. 
Also there is no systematics in the deviation as, for example,
that the collective coordinate approach would always underestimate the number
of bounces or that it would always predict too large a time interval 
during which bounces occur.

The results shown in the two bottom pictures of figure \ref{fig:XA} 
abandon all approximations. In that case the null-vector problem may 
occur. However, numerically this does not happen. For the cases displayed
in figure \ref{fig:XA} we always find that $X\ge0.1$. Of course, this value 
is correlated to $1/q$ that characterizes the regime of the short range repulsion.
Though the substitution $1\,\longrightarrow\,{\rm tanh}(qX)$ affects the 
coefficients $a_i(X)$ only in the moderate regime $|X|\lesssim3$, significant 
changes for $X(t)$ and $A(t)$ are measured. Yet, the number of changes as well 
as the interaction times vary with $v$ and $q$. Thus much of the predictive 
power of these calculations is lost.

\begin{figure}[t]
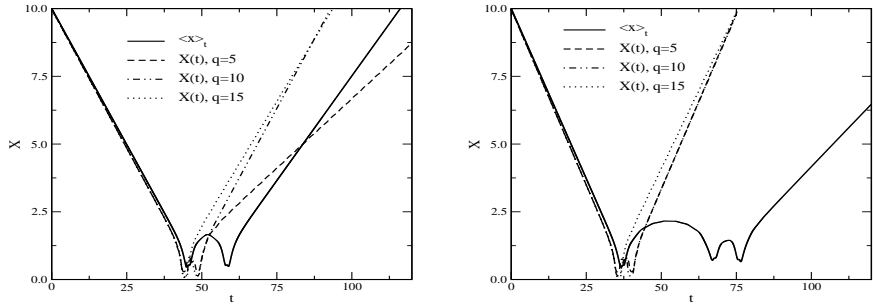

\bigskip
\centerline{
\epsfig{file=colq1.eps,width=5.4cm,height=4cm}\hspace{0.6cm}
\epsfig{file=colq2.eps,width=5.4cm,height=4cm}}
\caption{\label{fig:tuneQ}Effect of tuning $q$ in Eq.~(\ref{eq:cc4mod})
for two velocities, $v=0.201$ (left panel) and $v=0.251$ (right panel).}
\bigskip
\end{figure}

It has become apparent that the quality of the collective coordinate approach 
strongly depends on the applied approximations and no rigorous conclusion can be
drawn. Similar to Ref. \cite{Demirkaya:2017euk} one might consider the new 
parameter $q$ a tunable variable. Results for such simulations are compared to 
the mean value of the full solution in figure \ref{fig:tuneQ}.
While there are some variations with $q$, the main structure of several bounces 
over a long time interval cannot be adjusted.

\subsection{Extraction of critical velocities}

Once kink and antikink start to separate, the major share of the energy is stored in 
the translational motion. This raises expectations that the predictions for the critical 
velocities above which no bounces occur agree for the two approaches. However, this 
is only partially the case. The exact value for the critical velocity from the field 
equations of $0.260$ \cite{Moshir:1981ja,Campbell:1983xu} is underestimated within the 
collective coordinate approach of Eq.~(\ref{eq:cc4mod}) to be $0.204$ when $q=10$. However, 
that particular value again changes with $q$. For example, bounces are observed for 
$v=0.4$ when $q=5$.

We compare the initial and final velocities above critical velocities as 
extracted from $\frac{\partial\langle x\rangle_t}{\partial t}$ and 
$\dot{X}_\infty=\frac{\partial X(t)}{\partial t}$ as $t\to\infty$. Subsequently 
the collective coordinate velocity is written in relativistic kinematics
$v_f=\frac{\dot{X}_\infty}{\sqrt{1+\dot{X}_\infty^2}}$. Since there are
still oscillations of the shape mode on top of the translation those velocities
are not constant and the results listed in table \ref{tab:vel} are obtained
from averaging over numerous such oscillations.
\begin{table}[t]
\centerline{
\begin{tabular}{c|c|c|c|c}
&& \multicolumn{3}{|c}{coll. coord.}\cr
\hline
$v$ & PDE  & $q=5$ & $q=10$ & $q=15$\cr
\hline
0.4   & 0.279  & 0.216 & 0.399 & 0.340 \cr
0.5   & 0.390  & 0.447 & 0.496 & 0.404 \cr
0.6   & 0.494  & 0.569 & 0.577 & 0.388 \cr
0.7   & 0.595  & 0.700 & 0.612 & 0.699 \cr
0.8   & 0.697  & 0.818 & 0.806 & 0.808 \cr
0.9   & 0.797  & 0.897 & 0.879 & 0.906
\end{tabular}}
\caption{\label{tab:vel}Comparison of the predicted final velocities, $v_f$. The entry PDE 
refers to $\frac{\partial\langle x\rangle_t}{\partial t}\big|_\infty$ and the $q$ 
columns contain $\dot{X}_\infty$ originating from Eq.~(\ref{eq:cc4mod}).}
\end{table}
Again we see that the collective coordinate approach reproduces the exact
results only qualitatively and that particularities (here represented by
a strong $q$ dependence) matter. We also observe from the PDE entry of 
table \ref{tab:vel} that for large initial velocities the full field 
equations predict reduced final velocities. That is, even without bounces, 
some energy is stored in modes other than the translation.

\subsection{Mapping collective coordinates and solution to full field equation}
\label{sec:map}

By pure definition, Eq.~(\ref{eq:xn}) $\langle x\rangle_t$ is non-negative.
Thus a direct comparison with $X(t)$ may be misleading. We therefore attempt to rebuild the 
time dependent field from Eq.~(\ref{eq:cc4mod}) with $q=10$ by substituting $X(t)$ and 
$A(t)$ and then compare that configuration to the solution from the field equations, 
Eq.~(\ref{eq:PDE}) with the same initial velocity. 

\begin{figure}[b]
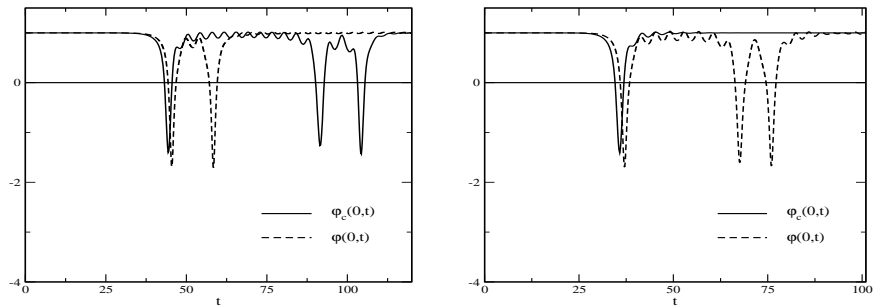

\bigskip
\centerline{
\epsfig{file=phi0201.eps,width=5.4cm,height=4cm}\hspace{0.6cm}
\epsfig{file=phi0251.eps,width=5.4cm,height=4cm}}
\caption{\label{fig:phi0}Time dependence of the field at the origin. Left panel: 
$v=0.201$, right panel: $v=0.251$. Collective coordinate results are without any 
approximation, {\it i.e.}, similarly to the bottom graphs of figure \ref{fig:XA}.}
\bigskip
\end{figure}

\begin{figure}[t]
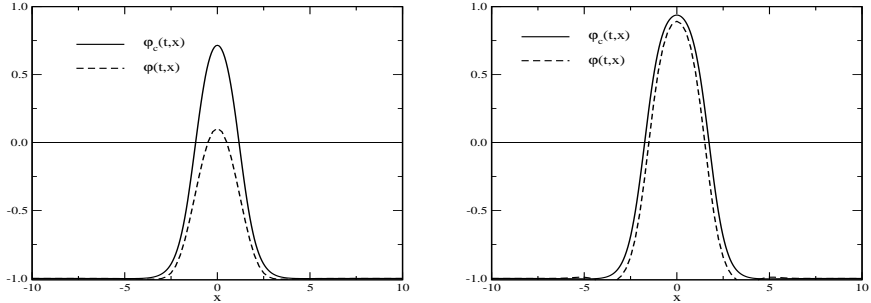

\bigskip
\centerline{
\epsfig{file=phit47v201.eps,width=5.4cm,height=4cm}\hspace{0.6cm}
\epsfig{file=phit50v201.eps,width=5.4cm,height=4cm}}
\caption{\label{fig:phit}Fields at different times for $v=0.201$.
Left panel: $t=47$, right panel: $t=50$.}
\bigskip
\end{figure}

We analyze the (dis)agreement in two ways. First, in figure \ref{fig:phi0} we consider 
the field at the center ($x=0$) as a function of time and, second, we contrast the 
configurations along the coordinate axis at different times in figure \ref{fig:phit}.
As in figure~\ref{fig:XA} (with $v=0.201$) we observe that the time between the
first two collisions of kink and antikink is overestimated by the collective 
coordinate description. We also see that the collective coordinate approach yields 
two nearby bounces at late times when the field equations predict well separated 
kink-antikink structures. Similarly the number and positions of bounces are not 
correct for $v=0.251$ either. Interestingly enough, the collective coordinate 
result for $v=0.201$ agrees in shape (though not with respect to the time scale) 
with the exact result of $v=0.251$. This again suggests that details of the 
parameterization matter and that the velocities should not be literally compared.

A further striking difference between the two approaches seen in figure~\ref{fig:phit} 
is that, at particular times, the peak amplitudes of the field configurations differ
significantly. Of course, this just reflects that at those particular times which
exhibit significant differences one approach produces a bounce but the other does not.

\section{Comparison: $\phi^6$ model}
\label{sec:phi6}

Without going into much detail we briefly reflect on related studies within
the $\phi^6$ model. It is interesting because bounce structures have been 
observed for the solutions to the field equations (analog to Eq.~(\ref{eq:PDE}))
\cite{Dorey:2011yw} even though no internal shape mode exists in the 
fluctuation spectrum of the soliton(s). This is a further indication that the 
shape mode is not the (only) explanation for bounces in kink-antikink scattering.

This model is defined by the (scaled) Lagrangian
\begin{equation}
\mathcal{L}_6=\frac{1}{2}\left[\dot{\phi}^2-\phi^{\prime2}\right]
-\frac{1}{2}\left(\phi^2+a^2\right)\left(\phi^2-1\right)^2\,,
\label{eq:deflag6}
\end{equation}
with the real parameter $a$. For $a\ne0$ there are two vacua $\phi_{\rm vac}=\pm1$
and the soliton solution mediates between them. We will only consider the $a=0$ case
when an additional vacuum, $\phi_{\rm vac}=0$ emerges. Then the soliton solutions
\begin{equation}
\phi_{K,\overline{K}}(x)=\left[1+{\rm exp}(\pm2x)\right]^{-\frac{1}{2}}
\label{eq:kink6}
\end{equation}
mediate between $0$ and $1$ (or $-1$ when changing the overall sign). The small amplitude 
bound state spectrum for both solitons only contains the translational zero mode. 

With two distinct soliton solutions available there are two independent initial 
conditions that relate to interactions between kink and antikink. They are usually referred 
to as kink-antikink ($K\overline{K}$) and antikink-kink ($\overline{K}K$) systems and 
the full field equations produce bounce structures for both systems \cite{Dorey:2011yw}.

\begin{figure}[b]
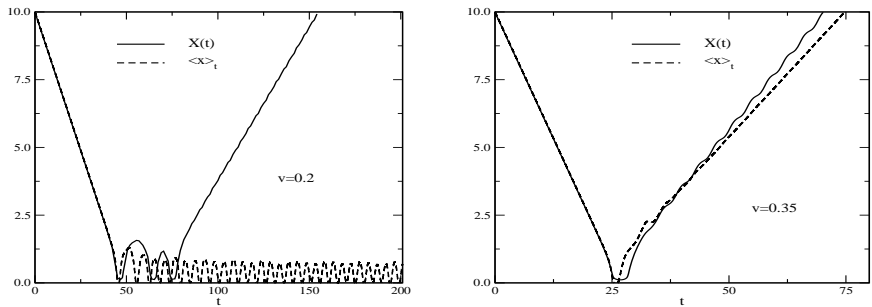

\bigskip
\centerline{
\epsfig{file=kkbarv20.eps,width=5.4cm,height=4cm}\hspace{0.6cm}
\epsfig{file=kkbarv35.eps,width=5.4cm,height=4cm}}
\caption{\label{fig:kkbar}Kink-antikink scattering in the $\phi^6$ model: comparison
of full and collective coordinate solutions for particular initial velocities.}
\bigskip
\end{figure}

\begin{figure}[b]
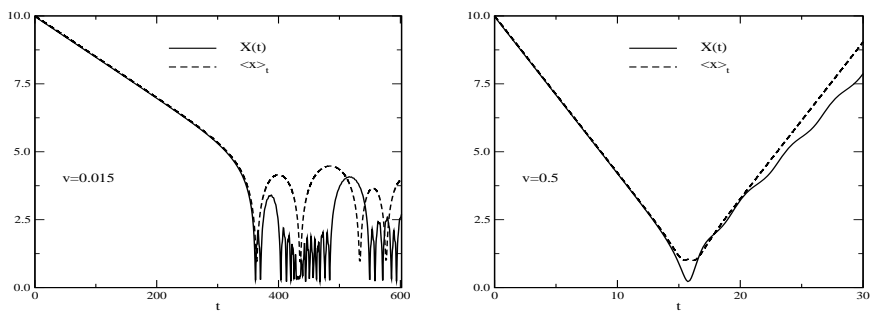

\bigskip
\centerline{
\epsfig{file=kbarkv015.eps,width=5.4cm,height=4cm}\hspace{0.6cm}
\epsfig{file=kbarkv50.eps,width=5.4cm,height=4cm}}
\caption{\label{fig:kbark}Same as Fig. \ref{fig:kkbar} for antikink-kink scattering 
in the $\phi^6$ model.}
\bigskip
\end{figure}

Though there is no shape mode in the spectrum of the small amplitude fluctuations its 
inclusion as in Eq.~(\ref{eq:ccSug}) may serve as working hypothesis for the collective 
coordinate approach to analyze the temporal storage of energy during the bounces. Furthermore
a Fourier analysis of $\phi(0,t)$, the solution of the field equations at the origin, in
the $\overline{K}K$ system shows large amplitudes at frequencies below threshold 
\cite{Dorey:2011yw}, suggesting that the energy is indeed stored in localized modes. Studies 
based on this hypothesis have been reported in Refs. \cite{Weigel:2013kwa,Takyi:2016tnc} and 
we reproduce typical results for the $K\overline{K}$ and $\overline{K}K$ systems in 
figures \ref{fig:kkbar} and \ref{fig:kbark}, respectively.  Again, the null-vector problem 
has been circumvented by approximating the coefficients of the terms quadratic in $A$ by 
their asymptotic values. As is by now standard, $\langle x\rangle_t$ has been extracted via 
Eq.~(\ref{eq:xn}) from the time dependent energy density deducted from Eq.~(\ref{eq:deflag6}) 
with $a=0$.
For $v\le0.289$ the $K\overline{K}$ is always trapped \cite{Dorey:2011yw}. Above that
velocity the two structures always reflect without any bounce. With the separation as the 
only collective coordinate the former feature can, by construction, not be reproduced. 
(See Ref.~\cite{Demirkaya:2017euk} for more details on this modification.) Calculations
with the shape mode added are shown in figure \ref{fig:kkbar}. Solutions with bounces are 
now produced for small enough velocities. In contrast to the full solution, however, trapping 
does typically not occur in the collective coordinate method. Neither is this accomplished 
by modifications similar to Eq.~(\ref{eq:cc4mod}) \cite{Takyi:2016tnc,Takyi:MSC2016} though 
the bounces are maintained by those modifications. On the other hand, for 
larger velocities the two calculations yield similar results.

In figure \ref{fig:kbark} the two methods are compared for the $\overline{K}K$ system
for which the critical velocity is much smaller: $v_{\rm c}=0.046$ \cite{Dorey:2011yw}.
There are obvious discrepancies below that velocity, while agreement is again observed
for larger values.

Interestingly enough despite all the obstacles observed, the ordering relations for 
the critical velocities
$$
v_{\rm c}(\phi^6,\overline{\rm K}{\rm K})\ll
v_{\rm c}(\phi^4,{\rm K}\overline{\rm K})
\lesssim v_{\rm c}(\phi^6,{\rm K}\overline{\rm K})
$$
in the different models are nevertheless observed in both the exact and the collective coordinate
approaches~\cite{Takyi:2016tnc}. The first relation is expected because those velocities differ
by an order of magnitude and even a coarse approximation should reproduce it. The second relation
may be a mere artifact.

\section{Conclusion and critical analysis}

The $\phi^4$ model in one space and one time dimensions has localized static solutions,
so-called (anti)kinks. Appropriate initial conditions for integrating the full field 
equations allow to simulate kink-antikink scattering as a prototype of particle interactions 
in the soliton picture. A number of interesting features emerge from these simulations as 
the relative initial velocity is varied. In particular, below a critical velocity
(multiple) bounces occur between kink and antikink.

The collective coordinate method has been introduced as (i) a manageable approximation to the 
full field equations to simplify partial to ordinary differential equations and (ii) a sensible 
means to identify important modes in the kink-antikink interaction. In particular the degrees
of freedom that are bound states of the single soliton have been considered. In the case 
of the $\phi^4$ model these are the translational zero mode and so-called shape mode.

Here we have confronted the collective coordinate method with the exact treatment of the 
field equations in the $\phi^4$ model. The collective coordinate approach reproduces well 
the first bounce observed in kink antikink collisions. This remarkably includes the 
acceleration shortly before kink and antikink sit on top of each other. However, when the 
separation is tiny, the non-linear equations of motion are particularly sensitive to small 
changes and the detailed pattern cannot be accommodated by the collective coordinate approach.

In fact, straightforward implementations of collective coordinates do not
produce acceptable agreement with the solutions of the field equations.
Intricate adjustments are needed to improve on the solutions. Even then 
significant discrepancies emerge. It is thus difficult to draw conclusions 
from the collective coordinate approach on the relevance of particular 
excitations, at least quantitatively. 

Qualitatively the collective coordinate approach to some extent supports the conjecture 
that energy storage in the shape mode excitation leads to bouncing kink-antikink
configurations. However, in that approach the amplitude of the shape mode is 
inflated compared to the exact solution. This suggests that other modes play a 
decisive role as well. It is also important to mention that within the interaction 
regime, {\it i.e.} when the distance between kink and antikink is not large, the 
shape mode ceases to be a solution to the fluctuation spectrum of $\varphi_{\rm cl}$. 
Though considering the linearized field equations is fully consistent only when 
the background is an extremum of the action - the kink-antikink system is not -
it may clarify whether the shape mode maintains its unique role from the single 
kink system. Such studies~\cite{ZLee,Graham:1998kz} show that the fluctuation 
spectrum of two (even widely) separated, stationary (anti)kinks differs significantly 
from that of a single kink. For fixed $X\gtrsim X_c\approx0.37$ the zero mode 
acquires negative frequency squared, indicating instability since two localized 
(anti)kinks at finite separation do not solve  the field equations. On the other 
hand, for $X\lesssim X_c$ the shape mode ceases 
to be bound which questions its incorporation as collective mode from the beginning. 
Also, the shape mode is a solution within the small amplitude approximation for 
deviations from a single kink. Smallness of this amplitude is not supported by the 
collective coordinate approach. (When computing scattering phase shifts about a single 
kink from the full non-linear field equations, amplitudes as small as $A\sim0.1$ yield 
results that deviate from the small amplitude approximation \cite{Abdelhady:2011tm}.) 
Hence most of the arguments leading to the incorporation of the shape mode are 
not rigorous in the regime of kink-antikink interactions. Another major issue 
in comparing the solutions of the two approaches is the fact that the relative 
velocities between the outgoing kink and antikink do not exactly match. This is 
most apparent from the different slopes of $X(t)$ and $\langle x\rangle_t$ in 
the above figures. 

In conclusion, we see that though the shape mode has its share in producing 
bouncing configurations in the kink-antikink scattering, the collective coordinate 
approach based on this mode is not sensible enough to reproduce the scattering process 
quantitatively. It is thus very likely that other modes are equally relevant for the 
temporary storage of energy. To clarify whether these are just a few modes one could
perform a thorough Fourier analysis in space and time of the difference
$$
\varphi(x,t)-\left[\varphi_{\overline{K}}(x-\langle x\rangle_t)+
\varphi_K(x+\langle x\rangle_t)-1\right]\,,
$$
with $\langle x\rangle_t$ computed via Eq.~(\ref{eq:xn}) from $\varphi(x,t)$, the
solution from the field equations. The result from that analysis may motivate 
more suitable collective coordinate parameterizations.

\section*{Acknowledgments}

The author gratefully acknowledges helpful contributions from A.M.H.H. Abdelhady and 
I. Takyi. This work is supported in part by the National Research Foundation of South 
Africa (NRF) by grant~109497.

\end{document}